\documentclass[runningheads,envcountsame]{llncs}

\usepackage{amsmath,amssymb}
\usepackage{listings}
\usepackage{xcolor}
\usepackage[most]{tcolorbox}
\usepackage{tikz}
\usetikzlibrary{fit,circuits.logic.US,positioning,calc}

\lstdefinelanguage{Coq}{
  morekeywords={Inductive,Definition,forall,Prop,Type,fun,match,with,end},
  sensitive=true,
  morecomment=[s]{(*}{*)},
  literate={¬}{{$\neg$}}1
}

\usepackage{hyperref}
\usepackage[capitalize]{cleveref}

\newcommand{\coqsymbol}{\begingroup\normalfont\includegraphics[height=1.7\fontcharht\font`\{]
{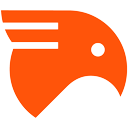}\endgroup}
\newcommand{\coqdocbaseurl}{https://vlamonster.github.io/RocqSAT/RocqSAT.}
\newcommand{\coqcorelibbaseurl}{https://rocq-prover.org/doc/V9.1.1/corelib/}
\newcommand{\coqstdlibbaseurl}{https://rocq-prover.org/doc/v9.0/stdlib/}
\newcommand{\coqident}[2]{\href{\coqdocbaseurl #1.html\##2}{\coqsymbol}}
\newcommand{\coqcorelibident}[2]{\href{\coqcorelibbaseurl#1.html\##2}{\coqsymbol}}
\newcommand{\coqstdlibident}[2]{\href{\coqstdlibbaseurl#1.html\##2}{\coqsymbol}}
\newcommand{\coqidenturl}[3]{\href{\coqdocbaseurl #1.html\##3}{\coqsymbol}}
\newcommand{\coqfile}[1]{\href{\coqdocbaseurl #1.html}{\nolinkurl{#1}}}

\makeatletter
\newcommand{\topbot}{\mathord{\mathpalette\top@bot\relax}}
\newcommand{\top@bot}[2]{\ooalign{$\m@th#1\top$\cr$\m@th#1\bot$\cr}}
\makeatother

\newcommand{\eval}[2]{[\![#2]\!]_{#1}}

\begin{document}

\title{Verification of a DPLL Transition System in Rocq}

\author{Julia Dijkstra\inst{1}\orcidID{0009-0000-6191-8534} \and
Benedikt Ahrens\inst{1}\orcidID{0000-0002-6786-4538}
}
\authorrunning{J. Dijkstra and B. Ahrens}

\institute{Delft University of Technology, Netherlands
\\
\email{juliadijkstra97@gmail.com}
\\
\email{B.P.Ahrens@tudelft.nl}
}

\maketitle

\begin{abstract}
We present a formal verification of an abstract transition-system presentation of the Davis-Putnam-Logemann-Loveland (DPLL) procedure in the Rocq proof assistant.
Following Nieuwenhuis et al., SAT solving is modeled as a set of rule-based transitions between states rather than as a concrete algorithm.
We formalize the syntax and semantics of propositional formulas, define the classical and base DPLL transition systems, and prove their key metatheoretic properties.
In particular, we establish correctness and completeness with respect to satisfiability, and we prove termination by showing that the transition relation is well-founded. The formalization extends the original abstract system by also including the pure literal rule.
Building on the verified transition system, we introduce an abstract notion of strategy and derive a terminating solver from any strategy satisfying suitable conditions.
We then implement a concrete strategy in Rocq and show that it satisfies the strategy specification.
\end{abstract}

\section{Introduction}
The Satisfiability (SAT) problem asks whether a Boolean formula admits an assignment of truth values to its variables that makes the formula true. Despite this simple formulation, SAT is a central problem in both theoretical computer science and practical computing. It was the first problem shown to be NP-complete \cite{karp2009reducibility}, and its generality makes it a natural target for encoding a wide range of combinatorial problems. As a result, improvements in SAT solving can often be translated into improvements for many other computational tasks.

SAT solvers are algorithms designed to decide this problem. Through sophisticated heuristics and search algorithms, they navigate enormous search spaces and make even very large SAT instances tractable \cite{Marques-Silva_Lynce_2014}. Beyond optimization and planning tasks, SAT solvers have become critical tools for verifying large mathematical proofs that are infeasible to check by hand, such as the Boolean Pythagorean Triples problem \cite{heule2016solving} resulting in a 200-terabyte proof. Another example is Schur Number Five \cite{heule2018schur}, producing a proof of roughly 2 petabytes.

These applications make it important to ensure the correctness of SAT solvers themselves since even small bugs could compromise the validity of the results they produce \cite{darbari2010industrial}. Interactive theorem provers (ITPs), such as Rocq \cite{rocq2025} or Isabelle \cite{nipkow2002isabelle}, provide a way to formally verify such systems and obtain strong guarantees about their behavior. In Rocq, definitions, theorem statements, and proofs are written in a formal language, and completed proofs are checked by a small trusted kernel. This means that the results established in this paper are justified not only by informal mathematical argument, but also by machine-checked proof terms.

The central objective of this paper is to formalize and verify, in Rocq, the abstract transition-system presentation of SAT solving introduced by Nieuwenhuis et al.~\cite{nieuwenhuis2006solving}. This presentation captures the Davis--Putnam--Logemann--Loveland (DPLL) procedure not as an executable algorithm, but as a collection of transitions between solver states. One contribution of this paper is to formalize these transition rules in Rocq and establish their key metatheoretic properties, namely correctness, completeness, and termination.
This objective is motivated by two considerations. First, the transition-system view gives a compact and transparent account of SAT solving, which makes it particularly suitable for formal reasoning. Second, it provides a natural starting point for later extensions toward more expressive frameworks such as DPLL(T), and therefore a useful basis for future verification efforts.

Closely related work has already been carried out in Isabelle/HOL. In particular, Blanchette et al.~\cite{blanchette2018verified} formalize abstract DPLL and CDCL calculi based on Nieuwenhuis et al.\ and refine them toward executable solvers.
Compared to that work, we contribute an abstract notion of \emph{strategy} for SAT solving, and instantiate this framework with a concrete strategy that is subsequently extracted to OCaml.

From this perspective, the aim of the paper is not to verify a highly optimized modern SAT solver directly, but rather to verify the correctness of an abstract model of SAT solving from which concrete solving procedures can later be derived. In doing so, the paper connects the abstract theory of DPLL with a machine-checked formalization in Rocq, and thereby provides a small but trustworthy verified core on which further extensions can build. Throughout the paper, references to formal objects are presented using hyperlink macros that allow the reader to move directly from the informal exposition to the corresponding definition, lemma, or theorem in the generated Rocq documentation.

The paper is organized as follows. Section~\ref{sec:propositional_logic} presents the necessary background on propositional logic and satisfiability. Section~\ref{sec:transition_systems} then introduces the transition system of Nieuwenhuis et al.~\cite{nieuwenhuis2006solving} and its formalization, including proofs of correctness, completeness, and termination. Building on this, Section~\ref{sec:solve_procedures} shows how solving procedures can be derived from the transition system using strategies and presents a concrete implementation of such a strategy. Finally, Section~\ref{sec:discussion} discusses the scope, limitations, and future directions of the formalization.

Summarizing the contributions of this work, we
\begin{itemize}
 \item prove an if-and-only-if characterization of satisfiability. Compared to Nieuwenhuis et al.~\cite{nieuwenhuis2006solving}, we not only show that derivations to a final non-fail state give a model, but also that, conversely, every satisfiable formula gives rise to such a derivation. For that converse direction we introduce a new normalization procedure.
 \item include the pure literal rule in our transition system. This rule is part of the original DPLL procedure, but Nieuwenhuis et al.~\cite{nieuwenhuis2006solving} leave it out of their transition system.
 To add this rule, we strengthen the entailment development, see \cref{def:entails,lem:entailment}.
 \item introduce an abstract notion of strategy, which lets one derive verified solvers from the transition system. We also give one concrete implementation of such a strategy, and extract and test it on a collection of SAT benchmarks.
 \item formalize the whole development in Rocq.
\end{itemize}
\noindent

For reproducibility, the full Rocq formalization is publicly available in the RocqSAT repository\footnote{\url{https://anonymous.4open.science/r/RocqSAT-2415}}, and HTML documentation is available on GitHub Pages\footnote{\url{https://vlamonster.github.io/RocqSAT/toc}}.
The formalization depends on Rocq \texttt{9.0}. In its current form, the development contains 187 lemmas, 7 theorems, 75 definitions, and 14 examples.
Of these 75 definitions, 30 are standard Rocq definitions, 36 are Equations definitions, and 9 are inductive Rocq definitions. Altogether, the formalization consists of 3555 lines of code, of which 2403 lines are proof scripts between \texttt{Proof.} and \texttt{Qed.}
Throughout the paper, definitions and results are accompanied by a link to the HTML documentation of the corresponding Rocq identifier (e.g., \coqident{Atom}{Atom}).

The repository also contains the extracted solver for Ocaml and several sample SAT problems on which to try the solver.

\section{Propositional Logic}
\label{sec:propositional_logic}

We begin by recalling the standard definition of satisfiability for propositional formulas. The notions introduced in this section are classical and can be found in standard textbooks on logic and formal reasoning, such as Huth and Ryan's \emph{Logic in Computer Science} \cite{huth2004logic}. We first introduce the basic syntactic elements used to construct propositional formulas.

For the following definitions we fix a finite set of propositional symbols $P$.

\begin{definition}[\coqident{Atom}{Atom}]
Elements $p \in P$ are called \textbf{atoms}.
\end{definition}

\begin{remark}
Since $P$ is finite, it is isomorphic to a finite subset of the natural numbers.
Consequently, we may identify atoms with elements of $\mathbb{N}$.
\end{remark}
A \textbf{literal} corresponds to either a positive or a negative occurrence of an atom,
enabling formulas to represent both assertions and their negations.

\begin{definition}[\coqident{Lit}{Lit}, \coqident{Neg}{neg}]
A \textbf{literal} over $P$ is either an atom $p \in P$ or its negation $\neg p$,
where negation is the involutive function that flips the sign.
\end{definition}
Literals can then be combined into \textbf{clauses}.

\begin{definition}[\coqident{Clause}{Clause}]
A \textbf{clause} is a finite disjunction of literals,
written $l_1 \vee \ldots \vee l_n$.
\end{definition}
Formulas are often expressed in \textbf{conjunctive normal form}. This form is convenient for reasoning about satisfiability and is the standard input for many SAT-solvers. Although this may appear to be a restriction, it does not lose generality, since every propositional formula can be transformed into an equivalent formula in conjunctive normal form.
\begin{definition}[\coqident{CNF}{CNF}]
A propositional \textbf{formula} in \textbf{conjunctive normal form} (CNF) is a finite conjunction of clauses,
written $c_1 \wedge \ldots \wedge c_k$.
\end{definition}
\begin{remark}
The Rocq formalization represents clauses and formulas as lists, without specifying an evaluation procedure.
The evaluation semantics are provided by the definition of \textbf{partial assignments} and \cref{def:ceval,def:leval,def:feval}.

\end{remark}
\begin{definition}
Let $f$ be a formula. A \textbf{partial assignment} $m$ for $f$ is a subset of literals appearing in $f$ such that
$\{p, \neg p\} \not\subseteq m$ for all propositional variables $p \in P$.
\end{definition}
While this set-based definition of partial assignments is natural, it is cumbersome for formal verification.
We instead represent partial assignments as lists.
This choice is motivated by the fact that the \texttt{list} type supports well-behaved induction principles, which facilitate formal reasoning, and also allows a partial assignment to be interpreted as a \textbf{trace} of assignments.

Each literal is also annotated with a type \texttt{Ann} \coqident{Evaluation}{Ann}.
We add the superscript $d$ to a literal, writing $l^d$, to indicate that it is a \textbf{decision literal}.
These annotations will be ignored during evaluation, but are essential for defining and reasoning about the DPLL transition rules.
\begin{definition}[\coqident{Evaluation}{PA}]
A \textbf{partial assignment} is a list of annotated literals.
\end{definition}
To recover set-like behavior, we require an additional property.
\begin{definition}
A partial assignment $m$ is \textbf{well-formed} \coqident{WellFormed}{WellFormed} in a formula $f$ if it is both \textbf{duplicate-free} \coqident{WellFormed}{NoDuplicates}, so that $\{p, \neg p\} \not\subseteq m$ for all $p \in P$, and \textbf{bounded} \coqident{WellFormed}{Bounded} by $f$, so that every literal in $m$ appears in $f$.
\end{definition}
Let $m$ be a partial assignment. We define a three-valued evaluation function
$$\eval{m}{\cdot} : \texttt{Lit} + \texttt{Clause} + \texttt{CNF} \;\to\; \{\texttt{T}, \texttt{F}, \texttt{U}\}.$$
In a slight abuse of notation, we write $l \in m$ to mean that $l$ appears in $m$ \emph{before} any occurrence of $\neg l$, evaluating the list from head to tail.
Note that this notation coincides with its standard definition for well-formed partial assignments.

\begin{definition}[\coqident{Evaluation}{l_eval}]
\label{def:leval}
For a literal $l$, we define
\[
\eval{m}{l} =
\begin{cases}
    \texttt{T} & \text{if } l \in m, \\
    \texttt{F} & \text{if } \neg l \in m, \\
    \texttt{U} & \text{otherwise.}
\end{cases}
\]
\end{definition}

\begin{definition}[{\normalfont\coqident{Evaluation}{c_eval}}]
\label{def:ceval}
For a clause $c$, we define
\[
\eval{m}{c} =
\begin{cases}
    \texttt{T} & \text{if some } l \in c \text{ satisfies } \eval{m}{l} = \texttt{T}, \\
    \texttt{F} & \text{if all } l \in c \text{ satisfy } \eval{m}{l} = \texttt{F}, \\
    \texttt{U} & \text{otherwise.}
\end{cases}
\]
\end{definition}

\begin{definition}[{\normalfont\coqident{Evaluation}{f_eval}}]
\label{def:feval}
For a CNF formula $f$, we define
\[
\eval{m}{f} =
\begin{cases}
    \texttt{T} & \text{if all } c \in f \text{ satisfy } \eval{m}{c} = \texttt{T}, \\
    \texttt{F} & \text{if some } c \in f \text{ satisfies } \eval{m}{c} = \texttt{F}, \\
    \texttt{U} & \text{otherwise.}
\end{cases}
\]
\end{definition}
One benefit of the list-based representation is that evaluation is defined for arbitrary lists of literals, not only for well-formed partial assignments. This lets us reason about more general lists throughout the formalization, while only introducing well-formedness assumptions in the lemmas that require them. By contrast, in a set-based approach, a set containing both $x$ and $\neg x$ cannot be evaluated without imposing an additional well-formedness condition.
\begin{example}
Consider the partial assignment $m = \neg x x$. This partial assignment is not well-formed, since it is not duplicate-free. However, it still provides a well-defined evaluation for all literals, with $\eval{m}{x} = \texttt{T}$ and $\eval{m}{\neg x} = \texttt{F}$.
\end{example}
\begin{definition}
A formula $f$ is \textbf{satisfiable} \coqident{Satisfiability}{Sat} if there exists a partial assignment $m$ that satisfies $f$.
Such an assignment is called a \textbf{model} \coqident{Satisfiability}{Model} of $f$, written $m \models f$. If no such assignment exists, $f$ is \textbf{unsatisfiable} \coqident{Satisfiability}{Unsat}.
\end{definition}

\section{Transition Systems}
\label{sec:transition_systems}
We now introduce the transition system framework used to describe the DPLL procedure.
\begin{definition}[\coqident{Trans}{State}]
A \textbf{state} is either the distinguished state \texttt{fail} or a pair of the form
$m \parallel f$, where $m$ is a partial assignment and $f$ is a formula.
\end{definition}
Having defined states, we now formalize how one state can evolve into another.
\begin{definition}
A \textbf{transition rule} $\Longrightarrow$ is a relation on states.
Given two states $s$ and $s'$, we say $s \Longrightarrow s'$ is a \textbf{transition} from $s$ to $s'$.
\end{definition}
To model the DPLL procedure, we consider a collection of transition rules.
In the following definitions, let $T$ denote such a set of transition rules.
\begin{definition}
A \textbf{transition system} $\Longrightarrow_T$ on $T$ is a relation on states,
where two states are related if one of the transition rules in $T$ applies.
\end{definition}
To state later lemmas and theorems, we consider sequences of transitions, motivating the reflexive–transitive and transitive closures of the transition system.
\begin{definition}[\coqident{Trans}{Derivation}]
Let $\Longrightarrow^*_T$ be the reflexive-transitive closure of $\Longrightarrow_T$.
Sequences of transitions under $\Longrightarrow^*_T$ are called \textbf{derivations}.
\end{definition}
\begin{definition}[\coqident{Trans}{DerivationStrict}]
Let $\Longrightarrow^+_T$ be the transitive closure of $\Longrightarrow_T$.
Sequences of transitions under $\Longrightarrow^+_T$ are called \textbf{strict derivations}.
\end{definition}
Finally, to reason about termination, we introduce the notion of a final state.
\begin{definition}[\coqident{Trans}{Final}, \coqident{Trans}{FinalB}]
A state is \textbf{final} in $T$ if there exists no state $s'$ such that $s \Longrightarrow_T s'$.
\end{definition}

\subsection{Transition System for DPLL}
We now specialize the general transition-system framework to the DPLL procedure. Following Nieuwenhuis et al.~\cite{nieuwenhuis2006solving}, we describe DPLL as a set of transition rules on states. We first give an informal overview of this transition-system view, before defining the rules formally.

In this formulation, the solver gradually extends a partial assignment until it either finds a satisfying assignment or reaches a conflict, applying any rule whose conditions are satisfied. The \emph{UnitPropagate} rule applies when a clause has become unit, meaning that all but one of its literals have been assigned false, so the remaining literal must be assigned true to satisfy the clause. The \emph{Pure} rule assigns a pure literal, that is, a literal whose negation does not occur in the formula, since doing so cannot create a conflict. The \emph{Decide} rule makes a branching choice by assigning a previously undefined literal as a decision literal. If a conflict is reached, then \emph{Backtrack} returns to an earlier decision point, undoes later assignments, and flips that decision, with the flipped literal no longer marked as a decision literal. If no earlier decision remains to be undone, \emph{Fail} applies, indicating that the formula is unsatisfiable.

This is formalized by the following transition rules:

{
\scriptsize
\begin{align*}
\textrm{UnitPropagate:}\quad
& m \parallel f \land (c \lor l) \;&\Longrightarrow&\quad m\, l \parallel f \land (c \lor l)
& \text{if }
& \begin{cases}
  c \text{ is false in } m \\
  l \text{ is undefined in } m
\end{cases}
\\
\textrm{Decide:}\quad
& m \parallel f \;&\Longrightarrow&\quad m\, l^{d} \parallel f
& \text{if }
& \begin{cases}
  l \text{ or } \neg l \text{ occurs in a clause of } f \\
  l \text{ is undefined in } m
\end{cases}
\\
\textrm{Fail:}\quad
& m \parallel f \land c \;&\Longrightarrow&\quad \texttt{fail}
& \text{if }
& \begin{cases}
  c \text{ is false in } m \\
  m \text{ contains no decision literals}
\end{cases}
\\
\textrm{Backtrack:}\quad
& m\, l^{d}\, n \parallel f \land c\;&\Longrightarrow&\quad m\, \neg l \parallel f \land c
& \text{if }
& \begin{cases}
  c \text{ is false in } m\, l^{d}\, n \\
  n \text{ contains no decision literals}
\end{cases}
\\
\textrm{Pure:}\quad
& m \parallel f \;&\Longrightarrow&\quad m\, l \parallel f
& \text{if }
& \begin{cases}
  l \text{ occurs in a clause of } f \\
  \neg l \text{ occurs in no clause of } f \\
  l \text{ is undefined in } m
\end{cases}
\end{align*}
}

We now distinguish between two transition systems. We refer to the transition system containing all of the rules above as the \textbf{classical DPLL} system. We also define the \textbf{base DPLL} system as the subsystem containing only \emph{Decide}, \emph{Fail}, and \emph{Backtrack}. We will later show that this smaller system is still sufficient for defining a complete solving procedure.
\begin{definition}[\coqident{Trans}{Trans}]
The \textbf{classical DPLL} system is a transition system on
$$C := \{\textrm{UnitPropagate}, \textrm{Decide}, \textrm{Fail}, \textrm{Backtrack}, \textrm{Pure}\}.$$
\end{definition}
\begin{definition}[\coqident{Trans}{TransB}]
The \textbf{base DPLL} system is a transition system on
$$B := \{\textrm{Decide}, \textrm{Fail}, \textrm{Backtrack}\}.$$
\end{definition}
\begin{remark}
Since most of our work concerns the classical DPLL system, the subscript $C$ is often omitted.
By slight abuse of notation, $\Longrightarrow$ is then used to refer both to the transition system and to its underlying transition rules.
\end{remark}
We now provide some examples of how this transition system can be applied.
\begin{example}
Consider the formula
$f = (x_1 \lor x_2) \land (x_1 \lor x_3) \land (x_2 \lor x_3) \land (\neg x_2 \lor \neg x_3)$.
In the following derivation, we hide any clauses that are already true under the current partial assignment and write them as ($\ldots$).
Then the transition rules can be applied as follows:
\begin{align*}
&&\emptyset \parallel (x_1 \lor x_2) \land (x_1 \lor x_3) \land (x_2 \lor x_3) \land (\neg x_2 \lor \neg x_3) \\
\qquad &\Longrightarrow &x_1 \parallel \ldots \land (x_2 \lor x_3) \land (\neg x_2 \lor \neg x_3)
&& (\text{Pure}) \\
\qquad &\Longrightarrow &x_1x_2^d \parallel \ldots \land (\neg x_2 \lor \neg x_3)
&& (\text{Decide}) \\
\qquad &\Longrightarrow &x_1x_2^d\neg x_3 \parallel \ldots
&& (\text{UnitPropagate})
\end{align*}
The first step applies \emph{Pure}, since $x_1$ occurs only positively in the formula. The second step applies \emph{Decide}, choosing $x_2$ as a decision literal. After this, the clause $(\neg x_2 \lor \neg x_3)$ becomes unit, so \emph{UnitPropagate} assigns $\neg x_3$. The resulting state is final, since every remaining clause is already satisfied and no rule applies. Since this final state is not \texttt{fail}, Theorem~\ref{theorem:final_model} implies that the partial assignment $x_1x_2^d\neg x_3$ is a model of $f$.
This can be easily verified.
\end{example}
\begin{example}
\label{example:rules_2}
Consider the formula
$f = x_1 \land (\neg x_1 \lor x_2) \land (\neg x_1 \lor \neg x_2)$.
Again, we replace clauses already true under the current partial assignment by ($\ldots$).
Then the transition rules can be applied as follows:
\begin{align*}
&&\emptyset \parallel x_1 \land (\neg x_1 \lor x_2) \land (\neg x_1 \lor \neg x_2) \\
\qquad &\Longrightarrow &x_1^d \parallel \ldots \land (\neg x_1 \lor x_2) \land (\neg x_1 \lor \neg x_2)
&& (\text{Decide}) \\
\qquad &\Longrightarrow &x_1^d x_2 \parallel \ldots \land (\neg x_1 \lor \neg x_2)
&& (\text{UnitPropagate}) \\
\qquad &\Longrightarrow &\neg x_1 \parallel x_1 \land \ldots
&& (\text{Backtrack}) \\
\qquad &\Longrightarrow &\texttt{fail}
&& (\text{Fail})
\end{align*}
The first step applies \emph{Decide}, choosing $x_1$ as a decision literal. After this, the clause $(\neg x_1 \lor x_2)$ becomes unit, so \emph{UnitPropagate} assigns $x_2$. This makes the clause $(\neg x_1 \lor \neg x_2)$ false, so \emph{Backtrack} undoes the assignments made after the last decision and flips that decision, yielding the state $\neg x_1 \parallel x_1 \land \ldots$. Since this state contains no decision literals and the clause $x_1$ is false, the \emph{Fail} rule applies. Since we derived \texttt{fail}, Theorem~\ref{theorem:final_unsat_refl} implies $f$ is unsatisfiable.
\end{example}

\subsection{Correctness}
We show that the classical DPLL transition system is sound and complete with respect to satisfiability.
In particular, we prove that the existence of a derivation from the initial state to a final state
precisely characterizes whether a formula is satisfiable or unsatisfiable.

We first show that every final state encodes a model of the formula.
\begin{theorem}[%
\coqident{Satisfiability}{final_model}]
\label{theorem:final_model}
Let $m$ be a partial assignment and $f$ a formula such that the state $m \parallel f$ is well-formed.
If $m \parallel f$ is a final state, then $m$ is a model of $f$.
\end{theorem}
Next, we show that if there exists a partial assignment $m$ satisfying $f$, then there exists a partial assignment $m'$ such that there is a derivation from the initial state $\emptyset \parallel f$ to a final state $m' \parallel f$. To establish this, we use a process called \textbf{normalization} \coqident{Normalization}{normalize}. This process has four steps. First, we \textbf{totalize} \coqident{Normalization}{f_totalize} the partial assignment, so that every literal occurring in $f$ is defined. Next, we \textbf{convert propagation annotations} \coqident{Normalization}{convert_prop} to decision annotations. We then \textbf{bound} \coqident{Normalization}{bound} the partial assignment by $f$, removing any literals whose atoms do not occur in $f$. Finally, we \textbf{deduplicate} \coqident{Normalization}{dedupe} the partial assignment, making it duplicate-free.
\begin{example}[normalization]
Consider the formula $f = (x_1 \lor x_2) \land x_3$ together with the partial assignment $m = \neg x_1x_1x_3^dx_4$.
We illustrate normalization by applying its steps to $m$.
\begin{align*}
\neg x_1x_1x_3^dx_4
&\longrightarrow \neg x_1x_1x_3^dx_4x_2 && (\text{totalize}) \\
&\longrightarrow \neg x_1^dx_1^dx_3^dx_4^dx_2^d && (\text{convert propagations}) \\
&\longrightarrow \neg x_1^dx_1^dx_3^dx_2^d && (\text{bound}) \\
&\longrightarrow x_1^dx_3^dx_2^d && (\text{dedupe})
\end{align*}
Since the original partial assignment $m$ is a model of $f$, this example also illustrates one of the invariants of the normalization procedure, namely that normalization is model preserving.
\end{example}
To show that the normalized state can be reached from the initial state and is final, we first establish the invariants of this normalization procedure. Each step establishes a new invariant that is preserved by all later steps, which gives rise to a number of intermediate lemmas. For brevity, we only state the results for the complete normalization procedure.

First, normalization preserves the property of being a model of $f$.
\begin{lemma}[%
\coqident{Normalization}{normalize_f}]
\label{lemma:normalize_f}
Let $m$ be a partial assignment and $f$ be a formula. If $m$ satisfies $f$, and $m'$ is obtained by normalizing $m$, then $m'$ also satisfies $f$.
\end{lemma}
After totalization, every literal occurring in $f$ is defined.
\begin{lemma}[%
\coqident{Normalization}{normalize_all_def}]
\label{lemma:normalize_all_def}
Let $m$ be a partial assignment and $f$ be a formula. If $m'$ is obtained by normalizing $m$, then every literal in $f$ is defined in $m'$.
\end{lemma}
After converting all annotations, every literal is a decision literal.
\begin{lemma}[%
\coqident{Normalization}{normalize_only_dec}]
\label{lemma:normalize_only_dec}
Let $m$ be a partial assignment. If $m'$ is obtained by normalizing $m$, then every literal in $m'$ is a decision literal.
\end{lemma}
After bounding, the partial assignment is bounded by $f$.
\begin{lemma}[%
\coqident{Normalization}{normalize_bounded}]
\label{lemma:normalize_bounded}
Let $m$ be a partial assignment and $f$ be a formula. If $m'$ is obtained by normalizing $m$, then $m'$ is bounded by $f$.
\end{lemma}
Finally, after deduplication, the partial assignment is duplicate-free.
\begin{lemma}[%
\coqident{Normalization}{normalize_no_duplicates}]
\label{lemma:normalize_no_duplicates}
Let $m$ be a partial assignment. If $m'$ is obtained by normalizing $m$, then $m'$ is duplicate-free.
\end{lemma}
It remains to show that a derivation to the normalized state exists. For this, we first prove an auxiliary lemma showing that every well-formed partial assignment consisting only of decision literals can be reached from the initial state $\emptyset \parallel f$.
\begin{lemma}[%
\coqident{Normalization}{normalize_derivation_aux}]
\label{lemma:normalize_derivation_aux}
Let $m$ be a partial assignment and $f$ be a formula. If $m$ is well-formed in $f$, and every literal in $m$ is a decision literal, then there exists a derivation from the initial state $\emptyset \parallel f$ to the state $m \parallel f$.
\end{lemma}
We obtain the desired result by specializing this lemma to the normalized partial assignment.
\begin{corollary}[%
\coqident{Normalization}{normalize_derivation}]
\label{corollary:normalize_derivation}
Let $m$ be a partial assignment and $f$ be a formula. If $m'$ is obtained by normalizing $m$, then there exists a derivation from the initial state $\emptyset \parallel f$ to the state $m' \parallel f$.
\end{corollary}
We now show that satisfiability is reflected in the existence of a derivation to a final state.
The forward direction follows directly from \cref{theorem:final_model}. The converse direction, however, does
not appear in this form in Nieuwenhuis et al.~\cite{nieuwenhuis2006solving}, and is proved here using the normalization
procedure introduced above.
\begin{theorem}[%
\coqident{Satisfiability}{final_sat_refl}]
\label{theorem:final_sat_refl}
Let $f$ be a formula. There exists a derivation from the initial state $\emptyset \parallel f$ to some final
state $m \parallel f$ if and only if $f$ is satisfiable.
\end{theorem}
There is an analogous theorem for unsatisfiability, but before stating and proving it, we first introduce the notions of \textbf{model-preserving literals} and \textbf{locally model-preserving literals}, which we will need in order to define \textbf{entailment}.
\begin{definition}
Let $f$ be a formula and $l$ a literal. We say that $l$ is \textbf{model-preserving} in $f$ if, for every partial assignment $m'$ satisfying $f$, the partial assignment $m'l$ satisfies $f$. For a partial assignment $m$, we say that $l$ is \textbf{locally model-preserving} in $f$ relative to $m$ if, for every partial assignment $m'$ satisfying both $m$ and $f$, the partial assignment $m'l$ satisfies $f$.
\end{definition}
We can now define \textbf{entailment}. Intuitively, this property expresses that propagated literals appearing in the tails of partial assignments derived by the transition rules are logical consequences of the literals that precede them.
\begin{definition}[\coqident{Entails}{Entails}]\label{def:entails}
Let $m$ be a partial assignment and $f$ a formula. We say that $f$~\textbf{entails}~$m$, or that $m$ is \textbf{entailed} by $f$, if one of the following holds:
\begin{enumerate}
    \item Every model of $f$ satisfies $m$, and $m$ contains no decision literals.
    \item The partial assignment $m$ can be written as $m' l^d n$, where every model satisfying $m' l^d$ satisfies $n$, the tail $n$ contains no decision literals, and $m'$ is entailed by $f$.
    \item The partial assignment $m$ can be written as $m' l n$, where $l$ is model-preserving in $f$, every model satisfying $m' l$ satisfies $n$, $n$ contains no decision literals, and $m'$ is entailed by $f$.
\end{enumerate}

\end{definition}
This definition is loosely based on the one given by Nieuwenhuis et al.~\cite{nieuwenhuis2006solving}, with the addition of the third case involving model-preserving literals. This extra case is needed to account for the \emph{Pure} rule in our transition system. One could go a step further and strengthen the classical \emph{Pure} rule by considering a literal pure whenever it appears with only one polarity among the clauses that remain undefined under a given partial assignment. In that setting, the third case would need to use locally model-preserving literals instead.

We now wish to relate entailment to unsatisfiability. For this, we show two things. Firstly, entailment is preserved by the transition system. Secondly, for partial assignments containing no decision literals, entailment coincides with logical entailment. We show the second first.
\begin{lemma}[%
\coqident{Entails}{entailment}]\label{lem:entailment}
Let $m$ and $m'$ be partial assignments and $f$ a formula. If $m$ contains no decisions, $m'$ is a model of $f$, and $f$ entails $m$, then $m' m$ is also a model of $f$.
\end{lemma}
\begin{remark}
The conclusion of the \texttt{entailment} lemma may seem unusual, since it gives no guarantee that $m'm$ is well-formed. This is not a problem for the results that follow. In fact, reasoning about such extensions will be useful in later proofs.
\end{remark}
The next lemma shows that entailment is preserved under the transition system. Since a full formal proof depends on a number of technical lemmas, we only give some intuition for the argument here. The proof proceeds by case analysis on the applied transition rule. For \emph{Fail}, the assumptions lead directly to a contradiction. For \emph{Decide} and \emph{Pure}, the second and third cases of the entailment definition apply directly. The argument for the \emph{Backtrack} and \emph{UnitPropagate} rules is more subtle, since they require unpacking how the original entailment was constructed. The key idea is that repeatedly unpacking this construction must eventually lead either to a contradiction or to a point where one of the entailment cases applies.
\begin{lemma}[%
\coqident{Entails}{trans_entails}]
Let $m$ and $m'$ be partial assignments and $f$ a formula. If there is a transition from $m \parallel f$ to $m' \parallel f$, and $f$ entails $m$, then $f$ also entails $m'$.
\end{lemma}
It follows immediately that the same holds not only for transitions, but also for derivations.
\begin{lemma}[%
\coqident{Entails}{derivation_entails}]
\label{lemma:derivation_entails}
Let $m$ and $m'$ be partial assignments and $f$ a formula. If there is a derivation from $m \parallel f$ to $m' \parallel f$, and $f$ entails $m$, then $f$ entails $m'$.
\end{lemma}
Before the main result, we first state one final lemma about derivations to \texttt{fail}.
\begin{lemma}[%
\coqident{Trans}{fail_predecessor}]
\label{lemma:fail_predecessor}
Let $m$ be a partial assignment and $f$ a formula. If there exists a derivation from $m \parallel f$ to \texttt{fail}, then there exists a partial assignment $m'$ such that there is a derivation from $m \parallel f$ to $m' \parallel f$, and a transition from $m' \parallel f$ to \texttt{fail}.
\end{lemma}
We now have all the tools needed to show how unsatisfiability is characterized by the \texttt{fail} state.
The reverse direction relies on a result we will show in the next subsection.
This theorem corresponds closely to part (1) of Theorem 2.12 in Nieuwenhuis et al.~\cite{nieuwenhuis2006solving}.
\begin{theorem}[%
\coqident{Satisfiability}{final_unsat_refl}]
\label{theorem:final_unsat_refl}
Let $f$ be a formula. There exists a derivation from the initial state $\emptyset \parallel f$ to the \texttt{fail} state
if and only if $f$ is unsatisfiable.
\end{theorem}
\begin{remark}
If one wishes to extend the transition system with \emph{Learn} and \emph{Forget} rules, the formula component of states may change. In that setting, a stronger notion of \textbf{equivalence} would be needed to prove this result.
\end{remark}

\subsection{Termination}
To show termination of any procedure based on the transition system introduced above, we use the standard notions of \textbf{accessibility} and \textbf{well-foundedness}. Intuitively, an element is accessible if all elements below it are themselves accessible, where the notion of below is determined by the relation under consideration. A relation is well-founded if every element is accessible, that is, if accessibility can always be established by a finite proof. This provides a convenient way to prove termination, since once the transition relation is shown to be well-founded, every sequence of transitions must eventually terminate. In the Rocq standard library, these notions are represented by the propositions \texttt{Acc} and \texttt{well\_founded}.
\begin{lstlisting}[language=Coq,caption={The \texttt{Acc} proposition from the Rocq standard library},label={lst:acc}]
Inductive Acc (x : A) : Prop :=
| Acc_intro : (forall y : A, R y x -> Acc y) -> Acc x.
\end{lstlisting}
\begin{lstlisting}[language=Coq,caption={The \texttt{well\_founded} proposition from the Rocq standard library},label={lst:well-founded}]
Definition well_founded := forall a : A, Acc a.
\end{lstlisting}

For the termination arguments below, we use several standard definitions and lemmas concerning relations from the standard library, including definitions of inclusion of relations, closure operations on relations,
and commutation of relations.
We now introduce a well-founded relation on bounded lists of natural numbers that will be used in the termination argument.
\begin{definition}[%
\coqident{Termination}{PrefixLt}]
Let $t$ be a natural number. We define $\texttt{PrefixLt}_t$ to be the relation on lists of natural numbers of length at most $t$.
We say that $m$ is smaller than $m'$ in the \textbf{prefix order} if $m$ and $m'$ agree on a common prefix and,
at the first position where they differ, the entry of $m$ is smaller than the corresponding entry of $m'$.
\end{definition}
Note that this is not quite the standard lexicographic order on lists. In the standard lexicographic order, a list can be smaller simply because it is shorter. By contrast, $\texttt{PrefixLt}_t$ only compares lists at the first position where they differ.

It follows from the well-foundedness of $<$ on the natural numbers that this relation is well-founded for every choice of $t$.
\begin{lemma}[%
\coqident{Termination}{wf_prefix_lt}]
For every $t \in \mathbb{N}$, $\texttt{PrefixLt}_t$ is well-founded.
\end{lemma}
We now formalize the termination argument used by Nieuwenhuis et al.\ to show that the classical DPLL system is well-founded. The idea is to equip states with two measures, \texttt{score} and \texttt{score\_total}, and to show that every transition rule decreases with respect to one of them. More precisely, we will prove that all transition rules except \emph{Fail} decrease with respect to the lexicographic product of these measures, together with the prefix order and the standard order $<$ on the natural numbers.

\begin{definition}[%
\coqident{Termination}{score}]
Let $f$ be a formula with $t$ distinct variables, and let $m$ be a well-formed partial assignment in $f$ of the form
$$m = m_0\, l_1^d\, m_1\, \dots\, l_n^d\, m_n,$$
where, for each $i$, the segment $m_i$ contains no decision literals.
The \texttt{score} of $m$ is the list
$$t - \texttt{length } m_0 \;::\; t - \texttt{length } m_1 \;::\; \dots \;::\; t - \texttt{length } m_n,$$
padded with additional occurrences of $t$ at the end until the resulting list has length $t$.
\end{definition}

\begin{definition}[%
\coqident{Termination}{score_total}]
Let $f$ be a formula with $t$ distinct variables, and let $m$ be a partial assignment. The \texttt{score\_total} of $m$ is $t - \texttt{length } m.$
\end{definition}
We briefly illustrate these measures on the transition rules from Example~\ref{example:rules_2}. For convenience, we write \texttt{TotalLt} \coqident{Termination}{TotalLt} for the standard order $<$ on the natural numbers. The key point is that along each transition, the measure \texttt{score} decreases with respect to \texttt{PrefixLt}, and if \texttt{score} remains unchanged, then \texttt{score\_total} decreases with respect to \texttt{TotalLt}.
\begin{example}
Consider the formula $f = x_1 \land (\neg x_1 \lor x_2) \land (\neg x_1 \lor \neg x_2)$.
This formula has two distinct variables, so $t = 2$.
Following the derived states from Example~\ref{example:rules_2}, the corresponding values of \texttt{score} and \texttt{score\_total} are:
\begin{align*}
&&\texttt{score}_f&\ \emptyset &= 2 :: 2, \qquad & \texttt{score\_total}_f\ \emptyset &= 2 \\
&&\texttt{score}_f&\ x_1^d &= 2 :: 2, \qquad & \texttt{score\_total}_f\ x_1^d &= 1 && (\text{Decide}) \\
&&\texttt{score}_f&\ x_1^d x_2 &= 2 :: 1, \qquad & \texttt{score\_total}_f\ x_1^d x_2 &= 0 && (\text{UnitPropagate})\\
&&\texttt{score}_f&\ \neg x_1 &= 1 :: 2, \qquad & \texttt{score\_total}_f\ \neg x_1 &= 1 && (\text{Backtrack})
\end{align*}
The \emph{Decide} step leaves \texttt{score} unchanged and decreases \texttt{score\_total}. By contrast, the \emph{UnitPropagate} and \emph{Backtrack} steps decrease \texttt{score} itself.
\end{example}
To complete the termination argument, it remains to account for the state \texttt{fail}. For this purpose, we introduce the relation \texttt{FailLt} \coqident{Termination}{FailLt}, which holds exactly when the left-hand side is \texttt{fail} and the right-hand side is a non-fail state.
In this way, \texttt{fail} becomes smaller than every non-fail state, which leads to the following relation on states.
\begin{definition}[%
\coqident{Termination}{StateLt}]
Let $f$ be a formula. We define the relation $\mathrel{\ll_f}$ on states by
$$
\ll_f \;=\; \texttt{FailLt} + (\texttt{PrefixLt}_{\texttt{score}},\ \texttt{TotalLt}_\texttt{score\_total}),
$$
using the union $+$ and the lexicographic product $(\_,\_)$ of relations.
Here, the bound for $\texttt{PrefixLt}_{\texttt{score}}$ given by the number of distinct atoms in $f$.
\end{definition}
It follows %
that this relation is well-founded.
\begin{lemma}[%
\coqident{Termination}{wf_state_lt}]
For every formula $f$, the relation $\ll_f$ is well-founded.
\end{lemma}
Finally, we show that $\Longrightarrow$ is well-founded. The key observation is that, for every state, there exists a formula $f$ such that the transition relation is included in $\ll_f$.
The claim then follows, using  %
the well-foundedness of $\ll_f$ established above.

\begin{theorem}[%
\coqident{Termination}{wf_trans}]
\label{theorem:wf_trans}
The relation $\Longrightarrow$ is well-founded.
\end{theorem}

\section{Solve Procedures}
\label{sec:solve_procedures}
The transition system from the previous section provides an abstract description of DPLL and establishes its key properties, in particular correctness, completeness, and termination. We now show how to turn this abstract system into a solving procedure. The main idea is to equip the transition system with a \emph{strategy} that determines, for each non-final state, which transition to take next. We then obtain a solver based on such a strategy by repeatedly applying it until a final state is reached.

\subsection{Abstract Solve Procedure}
We formalize this by introducing strategies as functions from states to optional successor states, where \texttt{None} indicates a final state, and that satisfy the following conditions.
\begin{definition}[\coqident{Strategy}{Strategy}]
A function $\texttt{next} : \texttt{State} \to \texttt{option State}$ is a \textbf{strategy} if it satisfies:
\begin{itemize}
    \item \texttt{next fail} $=$ \texttt{None}.
    \item For all states $s$, if \texttt{next} $s$ $=$ \texttt{None}, then $s$ is final in $B$.
    \item For all states $s$ and $s'$, if \texttt{next} $s$ $=$ \texttt{Some} $s'$, then $s \Longrightarrow^+ s'$.
\end{itemize}
\end{definition}

\noindent
We highlight two aspects of this definition that will be important in what follows.

First, we require only that a strategy step follows a strict derivation, rather than a single transition. This allows a strategy to combine several transition steps into one, for example by making multiple decisions at once. Although efficiency is not the main focus of this paper, this flexibility is still useful, since it allows the abstract notion of strategy to cover a broader range of solving procedures.

Second, the requirement that \texttt{next} returns \texttt{None} only on states that are final in $B$ is sufficient because the rules in $B$ already capture the essential search structure of DPLL. The additional rules \emph{UnitPropagate} and \emph{Pure} do not lead to fundamentally new search states. Rather, they only allow certain literals to be assigned by propagation instead of by explicit decision, which can speed up the search. Thus, being final in $B$ is equivalent to being final in $C$, as stated in the following lemma.
\begin{lemma}[%
\coqident{Trans}{final__final_b}]
\label{lemma:final__final_b}
A state is final in $C$ if and only if it is final in $B$.
\end{lemma}
Given a strategy \texttt{next}, we define the induced solver by iterating \texttt{next} from the initial state until a final state is reached. This solver is well-defined, since its iteration must terminate. Each application of \texttt{next} yields a step in the transitive closure $\Longrightarrow^+$ of the transition relation. Because $\Longrightarrow$ is well-founded by Theorem~\ref{theorem:wf_trans}, %
$\Longrightarrow^+$ is well-founded as well. Hence the iteration of \texttt{next} cannot produce an infinite sequence of states.
\begin{definition}[%
\coqident{Solve}{solve}]
Let \texttt{next} be a strategy. We define $\texttt{solve}(\texttt{next}) :\texttt{CNF}\to\texttt{State}$ to be the solver that maps a formula $f$ to the state obtained by repeatedly applying \texttt{next} to the initial state $\emptyset \parallel f$ until \texttt{next} returns \texttt{None}.
\end{definition}
By Lemma~\ref{lemma:final__final_b}, any state returned by \texttt{solve} is final. It therefore follows from Theorems~\ref{theorem:final_model} and \ref{theorem:final_unsat_refl} that the solver behaves as expected. If \texttt{solve} returns \texttt{fail}, then the formula $f$ is unsatisfiable, while if it returns a non-\texttt{fail} state of the form $m \parallel f$, then $m$ is a model of $f$.

\subsection{Concrete Solve Procedure}
We now instantiate the abstract notion of strategy by defining a concrete function \texttt{next\_state}. This function covers all transition rules of the classical DPLL system except for \emph{Pure}. It first checks whether the current state contains a conflict. If so, it determines whether there is a decision literal to backtrack to. If there is none, it applies \emph{Fail}. Otherwise, it applies \emph{Backtrack}. If no conflict is present, it tries to find a unit clause, and if one is found, it applies \emph{UnitPropagate}. Only if no conflict and no unit clause are found does it apply \emph{Decide}. In this way, \texttt{next\_state} yields a deterministic instance of the abstract strategy introduced above.

\Cref{lst:next_state} gives a simplified overview of this function, with the proof objects witnessing well-formedness omitted for readability. Here, \texttt{++p} denotes concatenation with a propagated literal, and \texttt{++d} denotes concatenation with a decision literal.
\begin{lstlisting}[
  language=Coq,
  caption={The concrete strategy \texttt{next\_state} \coqident{Strategy}{next_state}},
  label={lst:next_state},
  literate={¬}{{$\neg$}}1
]
Equations next_state (s: State): option State :=
next_state fail        := None;
next_state (state m f) :=
  match find_conflict m f with
  | Some c =>
    match split_last_decision m with
    | None         => Some fail
    | Some (m', l) => Some (state (m' ++p ¬l) f)
    end
  | _ =>
    match find_unit m f with
    | Some (c, l) => Some (state (m ++p l) f)
    | None        =>
      match find_decision m f with
      | Some l => Some (state (m ++d l) f)
      | None   => None
      end
    end
  end.
\end{lstlisting}
It remains to show that \texttt{next\_state} is indeed a strategy. For this, we verify two properties. First, \texttt{next\_state} is sound with respect to $\Longrightarrow$, meaning that whenever it returns a successor state, that successor is obtained by a single transition step, and hence by a strict derivation. Second, \texttt{next\_state} is complete with respect to $\Longrightarrow_B$, meaning that whenever a transition in the base DPLL system is possible, \texttt{next\_state} returns some successor state. Together with the fact that $\texttt{next\_state}\ \texttt{fail} = \texttt{None}$, this is enough to conclude that \texttt{next\_state} satisfies the definition of a strategy. These two properties are stated in the following lemmas.
\begin{lemma}[%
\coqident{Strategy}{next_state_sound}]
Let $s$ and $s'$ be states. If $\texttt{next\_state}\ s = \texttt{Some}\ s'$, then there is a transition from $s$ to $s'$.
\end{lemma}
\begin{lemma}[%
\coqident{Strategy}{next_state_exists}]
Let $s$ and $s'$ be states. If there is a transition from $s$ to $s'$ in the base DPLL system, then there exists a state $s''$ such that $\texttt{next\_state}\ s = \texttt{Some}\ s''$.
\end{lemma}

\subsection{Extraction to OCaml}

In addition to the formalization of the transition system and solving procedures, the concrete
solver was also extracted to OCaml using Rocq’s extraction mechanism. For this purpose,
the basic datatypes \lstinline!bool!, \lstinline!list!, and \lstinline!nat! were mapped to their OCaml counterparts using
the extraction directives shown in \cref{lst:extraction_ocaml}.

\begin{lstlisting}[language=Coq,
  caption={Extraction directives used for the OCaml solver},
  label={lst:extraction_ocaml}]
Extract Inductive bool => "bool" [ "true" "false" ].
Extract Inductive list => "list" [ "[]" "(::)" ].
Extract Inductive nat => "int"
[ "0" "(fun x -> x + 1)" ]
"(fun zero succ n -> if n=0 then zero () else succ (n-1))".
\end{lstlisting}
The extracted solver, together with a small runner and instructions for how to execute it, is
available in the repository.

To validate the extracted solver, we tested it on a
number of simple SAT instances, many of them taken from Gregory Duck’s collection of SAT
examples~\cite{duck}. For all instances on which the solver finished, it produced the correct result.
The largest example it was able to handle was \lstinline!zebra.cnf!, which has 155 variables and 1135
clauses. Of course, the size of a SAT instance does not necessarily correspond directly to its
difficulty, so this number should only be taken as a rough indication.
The current extracted solver does not scale well to larger instances. The main expected
reason is the choice of data structures in the formalization and in the concrete strategy.
Currently, many operations require traversing lists, which leads to linear-time access. A
more efficient implementation would use data structures supporting constant-time lookup,
such as arrays or similarly indexed data structures.
This would likely improve the practical
performance substantially.

\section{Discussion}
\label{sec:discussion}

In this paper, we formalized the DPLL transition system of Nieuwenhuis et al.~\cite{nieuwenhuis2006solving} in Rocq. The formalization covers the core correctness, completeness, and termination results for DPLL, and extends the original development by also formalizing the pure literal rule. Together, these results show that the transition-system presentation of DPLL can be mechanized in an interactive theorem prover while preserving the intended semantic guarantees of the abstract solver.

At the same time, the results of this paper should be understood as an abstract verified core for SAT solving. The formalization establishes correctness, completeness, and termination for the DPLL transition system, and derives both an abstract solver and a concrete strategy from it. This gives strong guarantees about the logical correctness of the procedure, but it does not yet amount to a verification of the behavior of modern industrial SAT solvers. In particular, the paper focuses on an abstract transition-system presentation and on a simple verified strategy, rather than on the highly optimized algorithms and data structures used in practice.

A first limitation is therefore the scope of the transition system itself. The formalization covers classical DPLL together with the pure literal rule, but it does not yet include extensions that are central to modern SAT solving, such as non-chronological backtracking, clause learning, clause forgetting, and restarts \cite{maric2010modernsat,blanchette2018verified}. These mechanisms are precisely what make modern CDCL-based solvers effective on large benchmark instances. As in Blanchette et al.~\cite{blanchette2018verified}, learn and backjump are best treated together, since combining them ensures that the solver continues to make progress after new clauses are learned. Relatedly, this paper does not yet cover the second part of Nieuwenhuis et al.~\cite{nieuwenhuis2006solving}, namely the extension from SAT to DPLL(T). The current formalization can therefore be seen as addressing the propositional core of their work, while leaving the SMT-oriented generalization for future work.

A second limitation concerns the concrete strategy derived from the transition system. Although it is formally verified as a strategy, it is intentionally simple and is not designed for performance. In particular, the function \texttt{next\_state} does not implement the \emph{Pure} rule. More generally, no serious experimental evaluation of the extracted solver has been carried out beyond testing correctness on small SAT instances. Thus, the current solver should be understood as a verified proof of concept rather than as a practically competitive SAT solver. Future work could improve this strategy by incorporating techniques used in practical SAT solvers, such as the two-watched-literal scheme for efficient unit propagation and the VSIDS branching heuristic for choosing decision literals \cite{moskewicz2001chaff}.

There are also limitations in the structure of the formalization itself. Currently, transition systems are treated as large relations rather than as modular collections of transition rules. This is sufficient for the results proved here, but it tends to produce large proofs that proceed by repeated case analysis over all rules at once. A more modular organization, for example in the style used by Mari\'c~\cite{maric2010modernsat}, could make the formalization easier to extend and maintain by splitting larger arguments into smaller lemmas for individual rules.

Finally, several natural extensions remain open. One possibility is to enrich the transition system with additional rules that bring it closer to modern SAT solvers, in particular backjumping, learning, forgetting, and restarts. Another possibility is to use the present formalization as a basis for a verified SMT solver, since the framework of Nieuwenhuis et al.~\cite{nieuwenhuis2006solving} is designed to generalize from SAT to DPLL(T). More abstract notions of solving procedure could also be considered, for example by letting strategies operate on states together with auxiliary context, allowing preprocessing steps and more efficient bookkeeping while still fitting into the same general framework.

\bibliography{references}

\appendix

\end{document}